\documentclass[aps,prb,twocolumn,showpacs,superscriptaddress,groupedaddress]{revtex4}  
\usepackage{graphicx}  
\usepackage{dcolumn}   
\usepackage{bm}        
\usepackage{amssymb}   

\begin{document}

\widetext


\title{Growth diagram and magnetic properties of hexagonal LuFe$_2$O$_4$ thin films}
\date{\today}

\author{Wenbin Wang}
\affiliation{Department of Physics, University of Tennessee, TN 37996, USA}
\affiliation{Materials Science and Technology Division, Oak Ridge National Laboratory, Oak Ridge, TN 37831, USA}

\author{Zheng Gai}
\affiliation{Materials Science and Technology Division, Oak Ridge National Laboratory, Oak Ridge, TN 37831, USA}
\affiliation{Center for Nanophase Materials Sciences, Oak Ridge National Laboratory, Oak Ridge, TN 37831, USA}

\author{Miaofang Chi}
\affiliation{Materials Science and Technology Division, Oak Ridge National Laboratory, Oak Ridge, TN 37831, USA}

\author{Jason D. Fowlkes}
\affiliation{Center for Nanophase Materials Sciences, Oak Ridge National Laboratory, Oak Ridge, TN 37831, USA}

\author{Jieyu Yi}
\affiliation{Department of Physics, University of Tennessee, TN 37996, USA}

\author{Leyi Zhu}
\affiliation{Materials Science Division, Argonne National Laboratory, Argonne, Illinois 60439, USA}

\author{ Xuemei Cheng}
\affiliation{Department of Physics, Bryn Mawr College, Bryn Mawr, PA 19010, USA}

\author{David J. Keavney}
\affiliation{Advanced Photon Source, Argonne National Laboratory, Argonne, Illinois 60439, USA}

\author{Paul C. Snijders}
\affiliation{Materials Science and Technology Division, Oak Ridge National Laboratory, Oak Ridge, TN 37831, USA}

\author{Thomas Z.  Ward}
\affiliation{Materials Science and Technology Division, Oak Ridge National Laboratory, Oak Ridge, TN 37831, USA}

\author{Jian Shen$^*$}
\affiliation{Department of Physics, University of Tennessee, TN 37996, USA}
\affiliation{Department of Physics, Fudan University, Shanghai 200433, China}

\author{Xiaoshan Xu$^*$}
\affiliation{Materials Science and Technology Division, Oak Ridge National Laboratory, Oak Ridge, TN 37831, USA}

\begin{abstract}
A growth diagram of Lu-Fe-O compounds on MgO (111) substrates using pulsed laser deposition is constructed based on extensive growth experiments. 
 The LuFe$_2$O$_4$ phase can only be grown in a small range of temperature and O$_2$ pressure conditions. 
 An understanding of the growth mechanism of Lu-Fe-O compound films is offered in terms of the thermochemistry at the surface. 
 Superparamagnetism is observed in LuFe$_2$O$_4$ film and is explained in terms of the effect of the impurity h-LuFeO$_3$ phase and structural defects .
\end{abstract}

\maketitle

\section{Introduction}
 Multiferroics have attracted great attention recently because of their promising new functionality and intriguing fundamental science. 
 A multiferroic material with a large ferroic polarization, high ordering temperature, and strong coupling between the ferroic orders is ideal for applications. 
 So far, those desired properties have not been realized in a single phase material. Multiferroics like BiFeO$_3$ where the magnetic and electric orders originate from different part of the structure have high ordering temperatures but weak coupling between different orders.\cite{Wang2003} 
 Other materials like TbMn$_2$O$_5$ exhibiting ferroelectricity due to the broken symmetry caused by the spiral magnetic moment have strong magneto-electric coupling.\cite{Hur2004} 
 However, here the ordering temperature is very low and the electric polarization is small. LuFe$_2$O$_4$ contains layers of Fe$_2$O$_2$ with a triangular lattice that are sandwiched by LuO$_2$ layers. 
 Combined with the mixed valance of Fe, the Fe$_2$O$_2$ layers in the triangular lattice form a charge ordered state at $T_{CO}$=320 K, followed by a ferrimagnetic order at $T_N$=240 K.\cite{Ikeda2005} 
 Significant changes in dielectric properties have been observed upon application of a small magnetic field at room temperature.\cite{Subramanian2006} 
 The relatively high transition temperature, large polarization, high magnetic coercivity and the strong magneto-electric coupling make LuFe$_2$O$_4$ a unique multiferroic material. 
 Recently, the possibility of fast switching and high tunability of LuFe$_2$O$_4$ due to the electronic origin of its charge order was demonstrated.\cite{Xu2011} 

Compared to the large amount of effort to study bulk LuFe$_2$O$_4$, there are only a couple of reported attempts to grow LuFe$_2$O$_4$ thin films on $\alpha$-Al$_2$O$_3$ (001) and on Si substrates using pulsed laser deposition (PLD).\cite{Liu2010, Rejman2011}  Liu et al found that the growth of LuFe$_2$O$_4$ on $\alpha$-Al$_2$O$_3$ (001) (with a target consisting of a sintered mixture of Lu$_2$O$_3$ and Fe$_2$O$_3$) needs substrate temperatures as high as 850 $^\circ$C.\cite{Liu2010} In addition, a significant deviation of the Lu:Fe stoichiometry from 1:2 was observed, which was attributed to different ablation efficiencies of Lu and Fe in the target. This problem was circumvented by enriching the Fe concentration of the target material. However, as a result, Fe$_3$O$_4$ and Fe$_2$O$_3$ impurities were introduced as intermediate layers between the LuFe$_2$O$_4$ film and the $\alpha$-Al$_2$O$_3$ substrate.
 
 In this paper, we present a comprehensive study on the growth of Lu-Fe-O compound thin films on MgO (111) substrates using pulsed laser deposition (PLD). 
 The experimentally constructed growth diagram shows that the parameter space for growing epitaxial LuFe$_2$O$_4$ thin films turns out to be a narrow window of temperature and O$_2$ pressure, which creates significant experimental difficulty. 
 Based on these results we have gained fundamental understanding of the growth of Lu-Fe-O compound films: the growth temperature needs to be high enough to stabilize the LuFe$_2$O$_4$ phase; on the other hand the loss of Fe at high temperature also produces phases other than LuFe$_2$O$_4$. These two effects cause narrow window of the growth condition producing LuFe$_2$O$_4$.
 Typical LuFe$_2$O$_4$ films appear to be superparamagnetic, which is consistent with the fact that the LuFe$_2$O$_4$ in the film is epitaxially sandwiched by an impurity phase of hexagonal LuFeO$_3$ (h-LuFeO$_3$).
 The current demonstration of epitaxial growth of LuFe$_2$O$_4$ thin films opens up new possibilities for studying multiferroicity of low dimensional LuFe$_2$O$_4$, tuning of its properties, and eventual functionalization.

 The paper is organized as the following: 
 Section II describes the experimental conditions used in this work; 
 Section III presents the experimental results including the growth diagram, structural characterizations and magnetism;
 Explanations of the observed growth diagram and magnetism of the films are proposed in Section IV. 

\section{Experimental conditions}

 Lu-Fe-O compound thin films were grown using PLD with a KrF ($\lambda$=248 nm) laser.
 The energy density of the laser is 2.5 J cm$^{-2}$ with a repetition rate of 1 Hz. 
 The target-substrate distance was 3.5 cm. 
 The thickness of the films grown in this study is approximately 100 nm. 
 The substrates are MgO (111) single crystals annealed in O$_2$ for 24 hours at 1100 $^\circ$C. 
 The target material used is polycrystalline LuFe$_2$O$_4$, whose properties are verified using powder X-ray diffraction (XRD) and a superconducting quantum interference device (SQUID). 
 After growth, the sample heating is turned off so that the sample cools to below 200 $^\circ$C at the same pressure as that of the growth condition within 5 minutes. 
 The substrates were clamped on a heater with a Pt foil in between. 
 The sample temperature was measured by a pyrometer using emissivity of 0.3. 
 In principle, all the parameters described above will have to be scanned and optimized in order to realize the growth of high quality LuFe$_2$O$_4$ thin films. 
 In this work, we are more focused on elucidating the mechanism of the growth. 
 Therefore, fine scans of the substrate temperature and the O$_2$ pressure were carried out to map out the growth diagram involving the growth of more than one hundred samples, while all the other parameters were kept constant. 
 
\begin{figure}[tb]
\centerline{
\includegraphics[width = 1\linewidth]{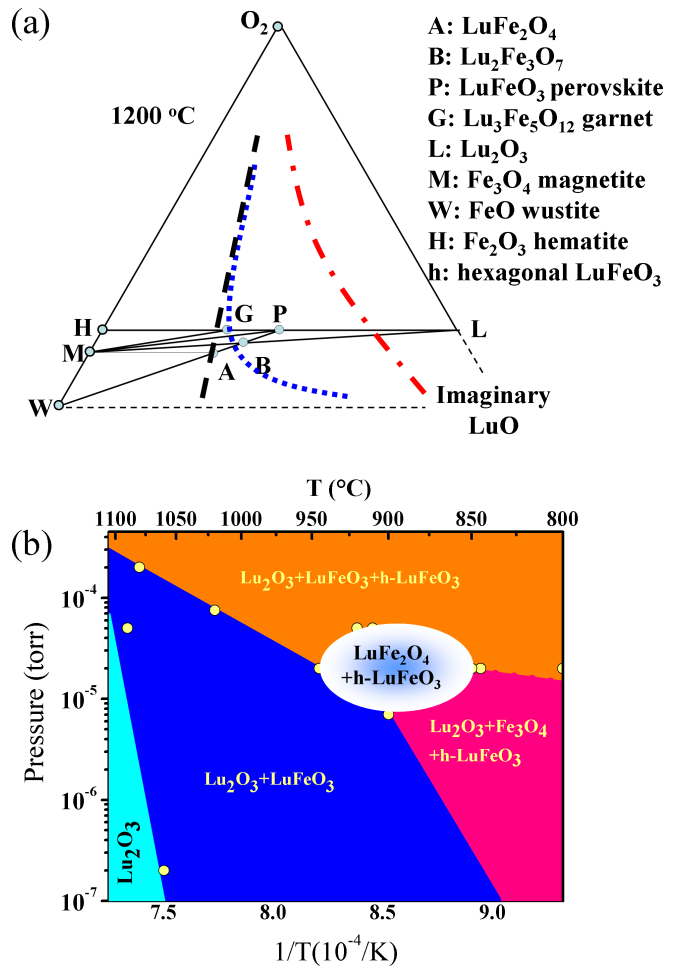}}
\caption{ \label{fig_phasegrowth} (Color online) 
 (a) A part of the phase diagram of the bulk Lu-Fe-O ternary system at 1200 $^\circ$C. 
 The thick dashed line represents the expected growth of the bulk Lu-Fe-O compound when the Lu:Fe stoichiometry is kept as 1:2 at 1200 $^\circ$C. \cite{Kumar2008, Sekine1976}
 The dash-dot line depicts the actual growth of Lu-Fe-O compound film at high temperature while the dotted line indicates the growth of at optimal temperature for LuFe$_2$O$_4$ film in this work.
 (b) The experimental growth diagram of the Lu-Fe-O compound thin films on MgO (111) substrates. The subset of data points that define the boundaries are shown as small circles.}
\end{figure}

\section{Results}

\subsection{Growth diagram}

In this work, we start from the ternary phase diagram of the bulk Lu-Fe-O system, a section of which is shown in Fig. \ref{fig_phasegrowth}(a) at 1200 $^\circ$C.\cite{Kumar2008, Sekine1976} This system belongs to the D-type of lanthanoid-Fe-O compounds for which there are four stable three-element phases: LuFe$_2$O$_4$ (A) and Lu$_2$Fe$_3$O$_7$ (B), LuFeO$_3$ (perovskite or P), and Lu$_3$Fe$_5$O$_{12}$ (garnet or G).\cite{Katsura1978} In principle, one way to form a single LuFe$_2$O$_4$ phase is to keep atomic ratio Lu:Fe=1:2 and vary the O$_2$ pressure, as shown as a thick dashed line in Fig. \ref{fig_phasegrowth}(a). 

\begin{figure}[tb]
\centerline{
\includegraphics[width = 1\linewidth]{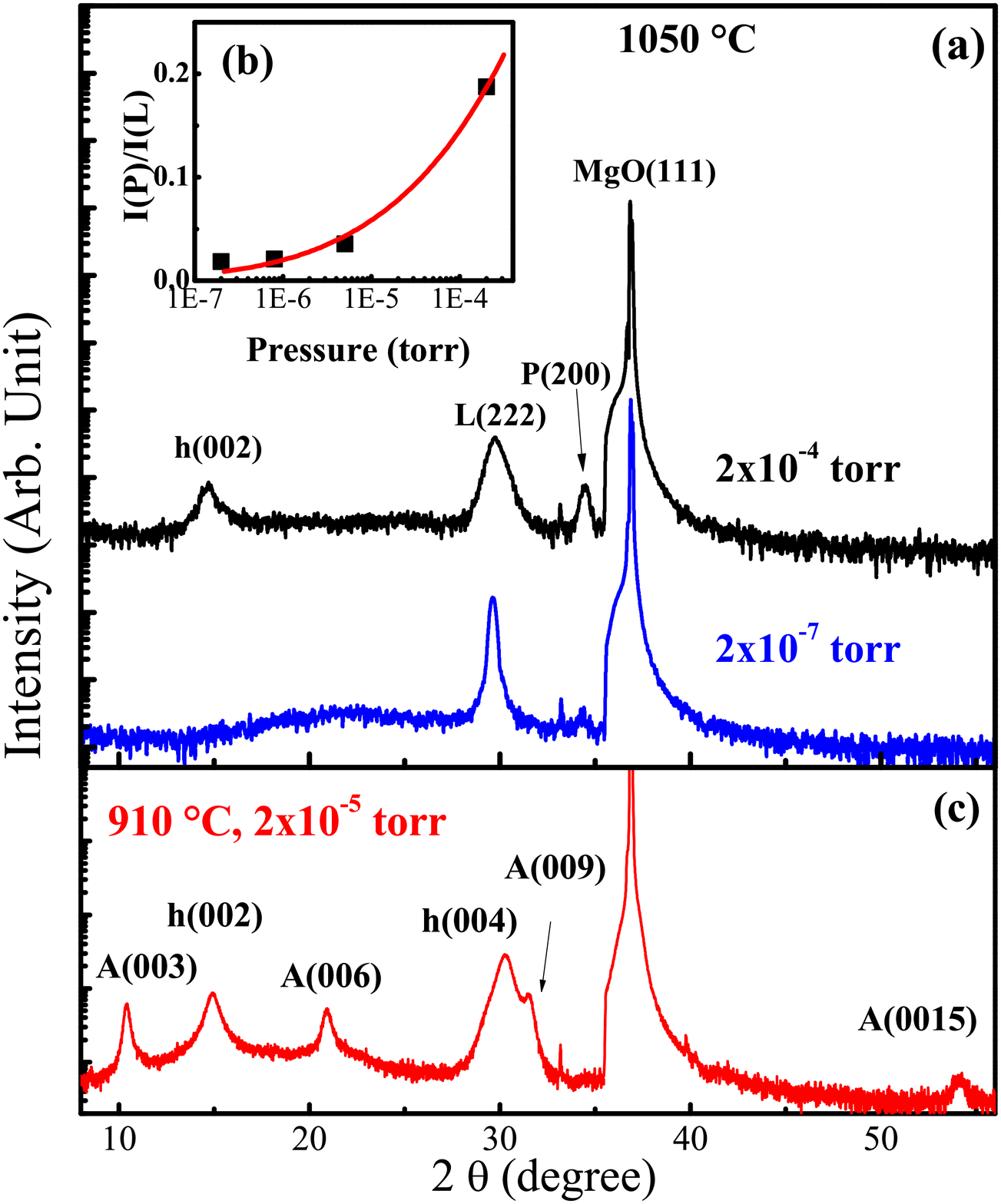}}
\caption{ \label{fig_XRD} (Color online) XRD data of the films grown at (a) $T$=1050 $^\circ$C and at two different O$_2$ pressures. (b) The ratio between the intensity of the P phase (LuFeO$_3$) I(P) and L phase (Lu$_2$O$_3$) I(L) as a function of the O$_2$ pressure. The line is a fit with the thermochemistry model (see text). (c) Typical XRD data of a LuFe$_2$O$_4$ film that shows the LuFe$_2$O$_4$ c-axis to be perpendicular to the substrate surface as expected. Also present is a h-LuFeO$_3$ as a impurity phase. }
\end{figure}

Figure \ref{fig_XRD}(a) presents the XRD data of films grown at 1050 $^\circ$C in various O$_2$ pressures. The LuFe$_2$O$_4$ phase is not observed. In addition, the Lu:Fe stoichiometry of the films is very different from that of the target. The dominant phase is always Lu$_2$O$_3$ (L). The concentration of LuFeO$_3$ rises with increasing O$_2$ pressure. At high enough O$_2$ pressure, h-LuFeO$_3$ compounds start to form.\cite{Kimizuka1983} 

In order to further elucidate the mechanism of the growth of Lu-Fe-O compound films, we carried out fine scans of the substrate temperature and the O$_2$ pressure to map out the growth diagram.
 Figure \ref{fig_phasegrowth}(b) is the resulting experimental growth diagram.
 The important observations can be summarized as follows:
 1) In the low temperature region the growth follows more or less the behavior predicted by the bulk phase diagram Fig. \ref{fig_phasegrowth}(a): at high pressure, the existing phases are LuFeO$_3$, Lu$_2$O$_3$ and h-LuFeO$_3$; when the pressure is decreased, the Fe$_3$O$_4$ phase starts to appear. This is consistent with the fact that LuFe$_2$O$_4$ and Lu$_2$Fe$_3$O$_7$ phases are not stable at low temperature.\cite{Kimizuka1983} 
 2) In the high temperature region, the growth deviates strongly from the thick dashed line in the bulk phase diagram Fig. \ref{fig_phasegrowth}
 (a) in that the Lu:Fe stoichiometry differs dramatically from that of the polycrystalline LuFe$_2$O$_4$ target. 
 The formation of Lu-Fe-O compounds in the films qualitatively follows the dash-dot line in Fig. \ref{fig_phasegrowth}(a).
 3) Only in the small range of pressure and temperature indicated by the elliptical area in Fig. \ref{fig_phasegrowth}(b), growth of LuFe$_2$O$_4$ is the most effective. In this case, the growth follows qualitatively the dotted line in Fig. \ref{fig_phasegrowth}(a). 
 Typical XRD data are diaplayed in Fig. \ref{fig_XRD}(c) showing both LuFe$_2$O$_4$ and h-LuFeO$_3$, indicating a deviation of Lu:Fe stoichiometry from that of the target even in this narrow window.\cite{Notephases}

\begin{figure}[tb]
\centerline{
\includegraphics[width = 1\linewidth]{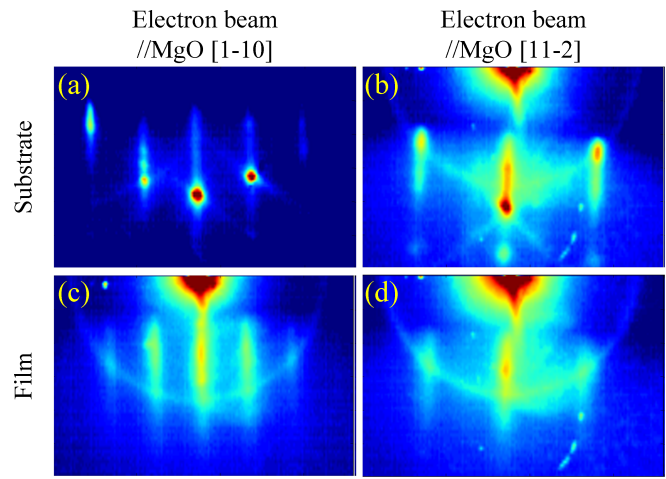}}
\caption{ \label{fig_RHEED} (Color online) RHEED images corresponding to the patterns of the MgO (111) substrate taken with the electron beam along [1-10] (a) and [11-2] (b) directions, and to the pattern of LuFe$_2$O$_4$ film taken with the electron beam along MgO [1-10] (c) and MgO [11-2] (d) directions. All panels have the same scale.}
\end{figure}

\begin{table}[t!]
\caption{Morphology and epitaxial relations of compounds grown on MgO (111) substrates.}
\label{tab:morph}\centering%
\begin{ruledtabular}
\begin{tabular}{c|c|c}
Compound&Morphology&Epitaxial relation\\ \hline
LuFe$_2$O$_4$ & Quasi 2D  &[001]//MgO [111], [100]//MgO [1-10] \\ \hline
Lu$_2$O$_3$ & Quasi 3D  &[111]//MgO [111], [1-10]//MgO [1-10] \\ \hline
LuFeO$_3$ & 3D &[100]//MgO [111], [001]//MgO [1-10] \\ \hline
Fe$_3$O$_4$ & 3D &[111]//MgO [111], [1-10]//MgO [1-10] \\ \hline
h-LuFeO$_3$ & Quasi 2D  &[001]//MgO [111], [1-10]//MgO [1-10]\\
\end{tabular}%
\end{ruledtabular}
\end{table}

\subsection{Structural characterization}

 The combination of in-situ structural characterization using Reflection High Energy Electron Diffraction (RHEED) and ex-situ characterization by XRD allows assignment of the epitaxial relation between the existing phases and the substrates. The results are given in Table I. 
 From the RHEED image, one can measure the in-plane lattice constant for the grown film. If three dimensional (3D) island growth occurs, the RHEED images correspond to the diffraction pattern of the transmitted electron beam which contains more structural information. Figure \ref{fig_RHEED} shows the RHEED images of the MgO (111) substrates and the LuFe$_2$O$_4$ films with the electron beams directed along MgO [1-10] or MgO [11-2]. The strong LuFe$_2$O$_4$ (003), (006) and (009) peaks observed in Fig. \ref{fig_XRD}(b) indicate that the epitaxial relation is LuFe$_2$O$_4$ [001]//MgO [111], which is expected because both faces have 3-fold rotational symmetry. The streaky RHEED patterns in Fig. \ref{fig_RHEED}(c) and (d) suggest quasi-2D growth of LuFe$_2$O$_4$. The in-plane lattice constants of the film can be calculated from the separation of the streaks calibrated by the RHEED pattern of the MgO substrates. It is consistent with the LuFe$_2$O$_4$ lattice constant 3.44 $\AA$ within the experimental uncertainty of 2\%. Hence, the in-plane epitaxial relation is LuFe$_2$O$_4$ [100]//MgO [1-10]. 
 This is unexpected from the point of view of lattice matching, which predicts LuFe$_2$O$_4$ [100]//MgO [11-2] because a $\sqrt{3}\times\sqrt{3}$ supercell of LuFe$_2$O$_4$ with 30 degree rotation along the [001] direction has less than 0.1\% mismatch with a 1$\times$1 of MgO (111) surface.\cite{supplementary} 
 Contrasting with the apparent quasi-2D growth of LuFe$_2$O$_4$, Lu$_2$O$_3$ forms quasi-3D structures on the substrate.\cite{supplementary} However, the RHEED pattern suggests a face centered cubic structure with a lattice constant half of that of bulk Lu$_2$O$_3$. The detailed structure is not clear at present. At low pressure and low temperature, the RHEED signal is dominated by the diffraction pattern of 3D Fe$_3$O$_4$ islands along the [11-2] direction, with the [111] direction perpendicular to the substrate surface. This is consistent with the XRD data. The lattice constants are the same as that of bulk Fe$_3$O$_4$ within the experimental uncertainty of 2\%.

\begin{figure}[tb]
\centerline{
\includegraphics[width = 1\linewidth]{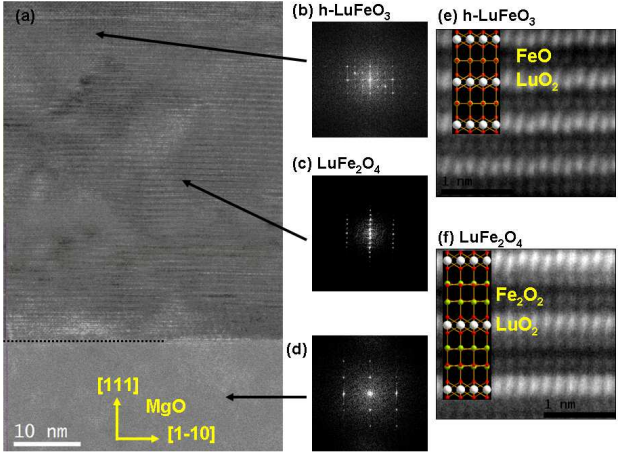}}
\caption{ \label{fig_TEM} (Color online) (a) Typical HRTEM image around the interface. (b-d) The Fourier transforms of various positions of (a), where (d) is from MgO substrate and (c) is from the majority of the film, and (b) is from the small top left part of the image. (e) and (f) are the atomic-resolution Z-contrast images corresponding to (b) and (c) respectively.}
\end{figure}

 High resolution transmission electron microscopy (HRTEM) reveals the detailed structure of the LuFe$_2$O$_4$ films. As shown in Fig. \ref{fig_TEM}(a), a layered structure of the film is obvious with some variation at different locations. The fast Fourier transforms (FFT) of the HRTEM image at different locations confirm the epitaxial relation observed from RHEED images: the FFT of the substrate (Fig. \ref{fig_TEM}(d)) indicates the reciprocal lattice of MgO viewed from [11-2] direction. The FFT of the majority of the film (Fig. \ref{fig_TEM}(c)) is consistent with the reciprocal lattice of LuFe$_2$O$_4$ viewed from the [1-10] direction, while at some locations (Fig. \ref{fig_TEM}(b)) it suggests h-LuFeO$_3$ viewed from the [100] direction. These two phases LuFe$_2$O$_4$ and h-LuFeO$_3$ were further confirmed by direct observation using atomic-resolution Z-contrast imaging, which is shown in Fig. \ref{fig_TEM}(e) and (f). The LuO$_2$-FeO-FeO-LuO$_2$ ordering in the LuFe$_2$O$_4$ phase and the LuO$_2$-FeO-LuO$_2$ ordering in the h-LuFeO$_3$ phase are clearly observed. 

 Although the intensity of the XRD peaks originating from the h-LuFeO$_3$ phase seems comparable to that of LuFe$_2$O$_4$ phase, the actual dominant phase is still LuFe$_2$O$_4$ due to the lower X-ray scattering cross section of the LuFe$_2$O$_4$ as compared with that of the h-LuFeO$_3$ phase.
 This is consistent with the low population of the h-LuFeO$_3$ phase in the HRTEM image. 
 In addition, the RHEED patterns of h-LuFeO$_3$ and LuFe$_2$O$_4$ are supposed to be different according to their structures.\cite{Qin2009, Yang2010, Magome2010} 
 The fact that the observed RHEED patterns do not show any indication of h-LuFeO$_3$ within the detection limit also suggests a low concentration of the h-LuFeO$_3$ phase in the films.

\begin{figure}[tb]
\centerline{
\includegraphics[width = 1\linewidth]{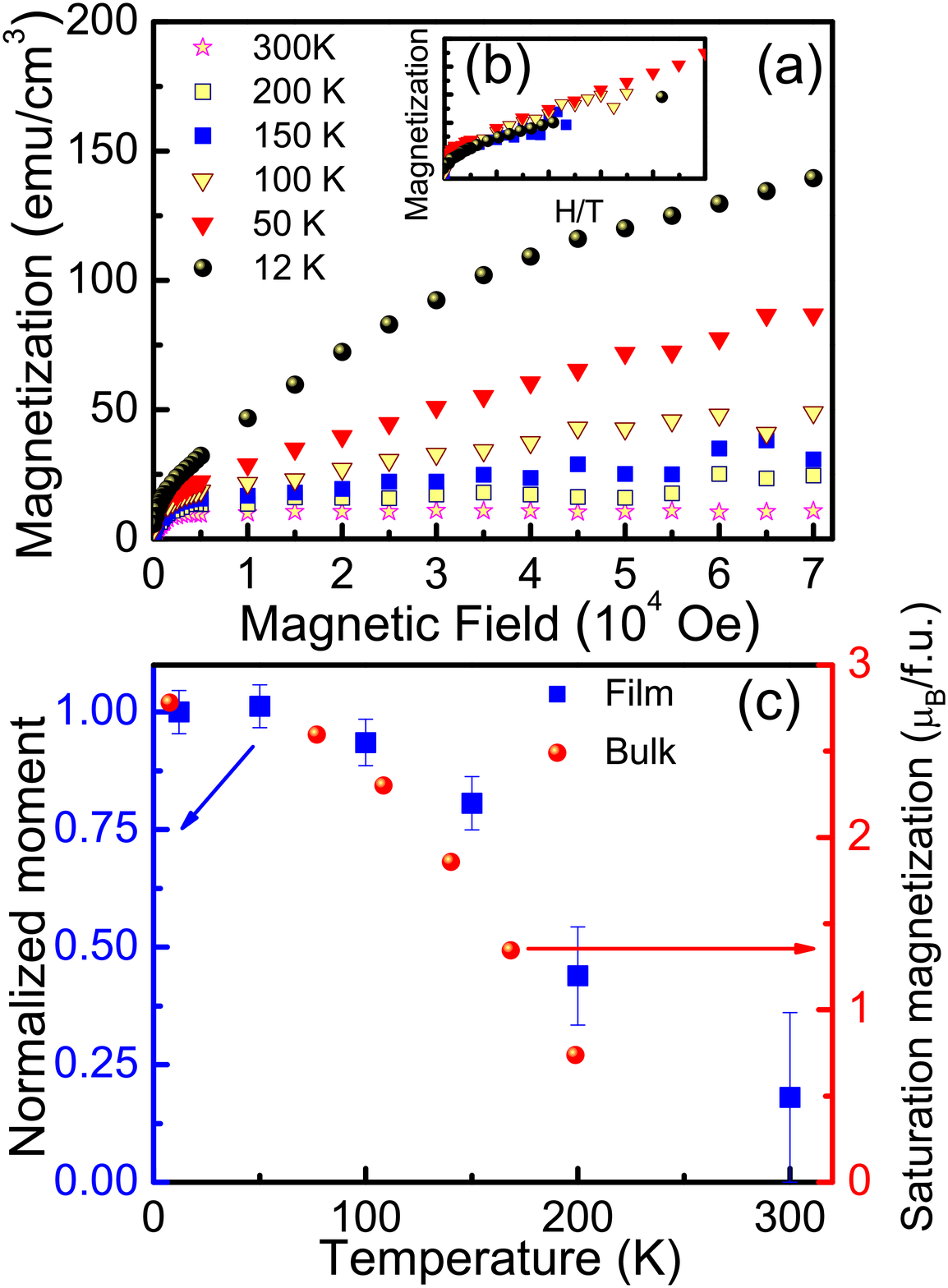}}
\caption{ \label{fig_mag4} (Color online) Magnetic properties of a typical LuFe$_2$O$_4$ film. (a) The field dependence of the magnetization at various temperatures. (b) Magnetization plotted again magnetic field over temperature ($H/T$). (c) The temperature dependence of the magnetic moment of the superparamagnetic phase normalized to the maximum value and the bulk saturation magnetization from Ref. [18]. The magnetic field is perpendicular to the plane of the film.}
\end{figure}

\subsection{Magnetic properties}
Ferrimagnetism, large magnetization and giant coercivity are of the key properties of LuFe$_2$O$_4$.\cite{Iida1987,Kiryukhin2008, Ko2009} This makes the study of the magnetic properties of LuFe$_2$O$_4$ films critical. As shown in Fig. \ref{fig_mag4}(a), little hysteresis is observed for these LuFe$_2$O$_4$ films, in contrast to the bulk.\cite{Iida1987} When magnetization is plotted against magnetic field over temperature ($H/T$), the data of 12, 50, 100 and 150 K fall on top of each other (Fig. \ref{fig_mag4}(b)), indicating superparamagnetic behavior.\cite{Notemag12K, Bean1959, Xu2008}

 Assuming superparamagnetism to be present, one can calculate the magnetic moment from the slope of the low field magnetization data with 
\begin{equation}
\frac{dM}{dH}
=
\frac{N\mu^2}{3kT}\mu_0,
\label{Eq_superpara}
\end{equation}

\noindent
where $\mu$, $N$, $\mu_0$ and $k$ are the moment of the superparamagnetic domains, number of the domains per unit volume, the vacuum permeability and the Boltzmann constant, respectively.\cite{Notefield, Bean1959, Xu2008} The magnetic moments normalized to their maximum value as a function of temperature are plotted in Fig. \ref{fig_mag4}(c), which follow the temperature dependence of the bulk saturation magnetization closely, suggesting that the Neel temperature of the films is not very different from the bulk value of 240 K. 

\section{Discussion}

\subsection{Growth diagram}

 The most surprising observation of the growth diagram is that the Lu-Fe-O compound formation at high temperature deviates strongly from the Lu:Fe stoichiometry of the target. Here we propose an explanation in terms of competition between nucleation and desorption of adatoms and its dependence on temperature and supersaturation.
 
 The residence time $\tau_{ad}$ of an adsorbed atom is given by:

\begin{equation}
\tau_{ad}
=\frac{1}{\nu}
\exp({\frac{E_{des}}{kT}}),
\label{Eq_tauad}
\end{equation}

\noindent
where $\nu$ is the vibrational frequency and $E_{des}$ is the desorption energy. Clearly, the residence time of an adatom is shorter at high temperature due to the higher desorption rate. The observed loss of Fe atoms suggests a smaller desorption energy (higher desorption rate) for Fe atoms. At low temperature, because $\exp({\frac{E_{des}}{kT}})$ is large for both Lu and Fe adatoms, the Lu:Fe stoichiometry can be close to that of the target.

The nucleation speed of deposited adatoms is:

\begin{equation}
J_{nuc}
\propto
(\frac{\Delta\mu^*}{T})^{1/2}
\exp(-\frac{\kappa}{\Delta\mu^*kT}),  
\label{Eq_Jnuc}
\end{equation}

\noindent
where $\Delta\mu^*$ is the effective supersaturation (molar bulk Gibbs free energy change with surface energy consideration), while $\kappa$ is proportional to the square of the edge energy of the nuclei per unit length.\cite{Markov1995} Therefore, at the high temperature limit, the nucleation speed decreases with temperature and a high supersaturation favors a high nucleation speed. Consider the reaction\cite{Note_reaction}

\begin{equation}
Fe+\frac{1}{2}Lu_2O_3+\frac{3}{4}O_2
\rightarrow
LuFeO_3,
\label{Eq_LFO113}
\end{equation}

\noindent
which takes place under thermodynamic equilibrium during the annealing time in between the laser pulses, the supersaturation of O$_2$ is related to the O$_2$ pressure as:

\begin{equation}
\Delta\mu^* _{O(ad)}
=
\Delta\mu^* _{0}(T)+\frac{3}{4}N_AkT ln(P_{O_2}),
\label{Eq_muO2}
\end{equation}

\noindent
where $N_A$ is the Avogadro constant.
Eq. (\ref{Eq_muO2}) suggests that higher O$_2$ pressure always corresponds to larger supersaturation, resulting in faster nucleation and better Lu:Fe stoichiometry. 
 
 Combining Eqs. (\ref{Eq_muO2}) and (\ref{Eq_Jnuc}), one has the analytical relation between the nucleation speed and the O$_2$ pressure:

\begin{eqnarray}
J_{nuc}
&\propto&[\frac{\Delta\mu^* _{0}(T)+\frac{3}{4}N_AkT ln(P_{O_2})}{T}]^{1/2} \nonumber\\
&&*\exp[-\frac{\kappa}{\Delta\mu^* _{0}(T)kT+\frac{3}{4}N_A ln(P_{O_2})(kT)^2}],
\label{Eq_JnucO2}
\end{eqnarray}

 Fig \ref{fig_XRD}(b) shows the XRD intensity (peak area) of the LuFeO$_3$ phase relative to Lu$_2$O$_3$ ($I(P)/I(L)$) as a function of the O$_2$ pressure at 1050 $^\circ$C.
 Assuming that the nucleation speed is proportional to the XRD intensity, one can fit experimental data with Eq. (\ref{Eq_JnucO2}).
 The result shows that $\Delta\mu^*(T=1050 ^\circ C)$ = 269 kJ mol$^{-1}$, similar to the bulk value found as $\Delta\mu_0 = \Delta H^0-T\Delta S^0=258.2$ J mol$^{-1}$, taking the $\Delta H^0$ = -41.8 kJ mol$^{-1}$ and $\Delta S^0$ = -121.4 J mol$^{-1}$ K$^{-1}$ and $T$ = 1050 $^\circ$C.\cite{Kumar2008a} 
 
 In the above analysis, the assumptions we made are: 1) the nucleation speed is proportional to the XRD intensity; 2) at high temperature the thermodynamic equilibrium gained during the annealing between the laser pulses determines the growth. These assumptions appear to be valid because the thermo-chemical parameters extracted from the model quantitatively agree with those from the literature. In other words, the growth of Lu-Fe-O at 1050 $^\circ$C can be described using equilibrium thermodynamics, presumably due to the thermodynamic equilibration that occurs in between the laser pulses. Here the competition between the desorption and nucleation determines the Lu:Fe stoichiometry. When the temperature is high enough, the time scales of the nucleation and desorption are comparable. In this case, change of nucleation speed (due to the change of supersaturation which is a function of O$_2$ pressure) affects the Lu:Fe stoichiometry dramatically.

 Based on this analysis, we expect the optimal growth conditions for LuFe$_2$O$_4$ films to be a narrow temperature and pressure window considering the necessary high temperature for the stability of LuFe$_2$O$_4$ phase that sets a lower limit, and the different desorption speed of Lu and Fe adatoms which sets an upper limit to the temperature. 
 As we have shown in Section III, this is indeed what has been observed in our experiments. 
 
\subsection{Magnetic properties}

 The observation of superparamagnetism in the LuFe$_2$O$_4$ films is unusual considering the bulk magnetic properties of LuFe$_2$O$_4$: an easy axis along the [001] direction with anisotropy energy as large as 100 K/spin and gigantic coercivity (9 T at 4 K).\cite{Iida1987, Kiryukhin2008, Ko2009,Phan2009,Park2009,Wang2009,Phan2010} 
 These unique bulk properties were attributed to the significant contribution of orbital magnetic moments (0.8 $\mu_B$/f.u.) plus the collective freezing of magnetic domains with the size of approximately 100 nm in the Fe$_2$O$_2$ layer and 30 nm along the [001] direction. \cite{Notefield, Iida1987}  
 The following scenario may explain the reduction of coercivity qualitatively: the structure of LuFe$_2$O$_4$ and h-LuFeO$_3$ both consist of layers of triangular lattice that can be epitaxial to each other nicely. 
 For LuFe$_2$O$_4$, the stacking is Fe$_2$O$_2$/LuO$_2$ while for h-LuFeO$_3$, FeO layers replace Fe$_2$O$_2$ layers. \cite{Qin2009, Yang2010} 
 From XRD data, one can see the co-existence of both LuFe$_2$O$_4$ and h-LuFeO$_3$ phases. 
 HRTEM indicates that the LuFe$_2$O$_4$ layers are divided into clusters (much smaller than the magnetic domain size in bulk) by the h-LuFeO$_3$ layers and defects. 
 According to a recent study, h-LuFeO$_3$ is weakly ferromagnetic, i.e. much less magnetic than LuFe$_2$O$_4$ \cite{Akbashev2011,LFO113Mag} 
 Therefore, when these LuFe$_2$O$_4$ clusters are much smaller than the dimensions of the magnetic domain  in the bulk, one expects to see a reduction in coercivity. 
 On the other hand, given the large anisotropy energy 100 K/spin, the observed hysteresis is too small even for clusters having a size as small as a few nanometers. Further study on the microscopic magnetic structure is needed to understand the difference between the bulk and films.

\section{Conclusion}
In conclusion, we studied the growth dynamics of LuFe$_2$O$_4$ films on MgO (111) substrates and constructed the growth diagram. 
 According to our understanding, application of the correct thermochemistry is the key to preferential formation of the LuFe$_2$O$_4$ phase: 
 1) at low temperature, LuFe$_2$O$_4$ is not a thermodynamically stable phase; 
 2) at high temperature, the Lu:Fe stoichiometry is off by so much due to the faster desorption of Fe adatoms that LuFe$_2$O$_4$ can not be formed;
 3) in a narrow range of substrate temperature and O$_2$ pressure, LuFe$_2$O$_4$ dominates the grown phases with some h-LuFeO$_3$ phase epitaxially sandwiched in between due to the loss of Fe atoms.
 Superparamagnetism is observed in the film of LuFe$_2$O$_4$ containing h-LuFeO$_3$ impurities. The extracted Neel transition temperature is similar to that of bulk. 
 
 This work reveals the growth mechanism of Lu-Fe-O compound thin films,
 paving the way to the growth of high quality LuFe$_2$O$_4$ thin films and offers an approach to tuning their properties. This will be critical for future applications using LuFe$_2$O$_4$, a unique multiferroic material with large polarizations, high ordering temperatures, and strong magneto-electric coupling.

 Research supported by the U.S. Department of Energy, Basic Energy Sciences, Materials Sciences and Engineering Division and performed in part at the Center for Nanophase Materials Sciences (CNMS) (Z.G., J.D.F) and ORNL's Shared Research Equipment (SHaRE)(M.C.) User Facility, which are sponsored at Oak Ridge National Laboratory by the Office of Basic Energy Sciences, U.S. Department of Energy.
 X.S. Xu acknowledges his research performed as a Eugene P. Wigner Fellow and staff member at the ORNL, managed by UT-Battelle, LLC, for the U.S. DOE under Contract DE-AC05-00OR22725.
 We also acknowledge partial funding supports from the National Basic Research Program of China (973 Program) under the grant No. 2011CB921801 (J.S.) and the US DOE Office of Basic Energy Sciences, the US DOE grant DE-SC0002136 (W.B.W).


$^*$ To whom correspondence should be addressed: xux2@ornl.gov, shenj5494@fudan.edu.cn.


\begin{references}
\bibitem{Notephases} Several phases do not show up in the XRD data including FeO and Lu$_3$Fe$_5$O$_{12}$. The fact that there are also no RHEED patterns related to these phases suggests that their population is minimal if not zero. The absence of those phases and the presence of the h-LuFeO$_3$ which is unstable in bulk may be related to the epitaxial preference.
\bibitem{Notemag12K} At 12 K, the remanence is 7.5 emu cm$^{-3}$ and the coercive field is approximately 300 Oe. This can be attributed to the existence of a trace amount of Fe$_3$O$_4$ islands because their temperature dependences are very similar to those of the samples with larger Fe$_3$O$_4$ populations (data not shown) and much larger remanence.
\bibitem{Notefield} Here the field should be high enough to saturate the impurity moment ($>$300 Oe).
\bibitem{Note_reaction} This analysis is only sensible at high temperature where desorption is significant. 
 At high temperature, there are basically only two reactions left: Eq. (\ref{Eq_LFO113}) and $Fe+\frac{1}{2}Lu_2O_3+\frac{3}{4}O_2\rightarrow h-LuFeO_3,$ (i.e. other Lu-Fe-O phases are not thermodynamically stable at high temperatures). 
 Unfortunately the thermo-chemical properties of h-LuFeO$_3$ are not available. 
 Therefore, only the analysis based on Eq. (\ref{Eq_LFO113}) can give us a degree of understandingof the growth.
\bibitem{LFO113Mag} Our ongoing work on pure h-LuFeO$_3$ films is also consistent with weak ferromagnetism.
\end{references}
\end{document}